\documentclass[11pt,a4paper]{article}

\usepackage{amsmath,amssymb}
\usepackage{epsfig,graphicx}
\usepackage{subfigure}
\usepackage{graphicx}
\usepackage{rotating}
\usepackage{cancel}
\usepackage{bm}
\usepackage{color}
\usepackage{comment}
\usepackage{cite}
\usepackage{psfrag}
\usepackage{slashed}
\usepackage{soul}
\usepackage{tikz}
\usepackage[normalem]{ulem}
\usepackage[export]{adjustbox}
\usepackage{hyperref}
\usepackage[mode=buildnew]{standalone}

\hypersetup{colorlinks=true,linkcolor=blue,citecolor=blue,urlcolor=blue}

\newtheorem{definition}{Definition}
\newtheorem{theorem}{Theorem}

\newcommand{\phip}{\varphi}

\renewcommand\[{\left[}

\newcommand{\exclude}[1]{}

\def\beq{\begin{equation}}
\def\eeq{\end{equation}}

\begin{document}
\numberwithin{equation}{section}
\title{
\Large{\textbf{Field Redefinitions in Classical Field Theory \\
with some Quantum
Perspectives}}}

\author{Juan~Carlos Criado$^{1}$, Joerg Jaeckel$^{2}$ and Michael Spannowsky$^{3}$\\[10pt]
\small{\em $^1$Departamento de F\'isica Te\'orica y del Cosmos, Universidad de Granada, } \\ \small{\em Campus de Fuentenueva, E–18071 Granada, Spain}\\[5pt]
\small{\em $^2$Institut f\"ur theoretische Physik, Universit\"at Heidelberg,} \\
\small{\em Philosophenweg 16, 69120 Heidelberg, Germany}\\[5pt]
\small{\em $^3$Institute for Particle Physics Phenomenology, Department of Physics,} \\
\small{\em South Road, Durham DH1 3LE, United Kingdom}\\[10pt]
}


\date{}
\maketitle

\begin{abstract}
\noindent

In quantum field theories, field redefinitions are often employed to remove redundant operators in the Lagrangian, making calculations simpler and physics more evident. 
This technique requires some care regarding, among other things, the choice of observables, the range of applicability, and the appearance and disappearance of solutions of the equations of motion (EOM). 
Many of these issues can already be studied at the classical level, which is the focus of this work. 
We highlight the importance of selecting appropriate observables and initial/boundary conditions to ensure the physical invariance of solutions. A classical analogue to the Lehmann-Symanzik-Zimmermann (LSZ) formula is presented, confirming that some observables remain independent of field variables without tracking redefinitions. Additionally, we address, with an example, the limitations of non-invertible field redefinitions, particularly with non-perturbative objects like solitons, and discuss their implications for classical and quantum field theories.

\end{abstract}

\newpage

\tableofcontents
\newpage


\section{Introduction}\label{sec:intro}
The standard way to set up a model of physics is to specify its action. The path integral then provides the implementation as a quantum theory. 
One of the advantages of this formalism is that one can freely choose the coordinates.
In the context of field theory, this means that one can do arbitrary field redefinitions.

For practical purposes, it is useful to specify the complete model with as few parameters as possible. 
Ideally, the terms in the Lagrangian should also have a reasonably simple physical interpretation. 
This is particularly relevant in the case of effective field theories, which often feature a large number of parameters and potentially complicated higher-dimensional operators.
An important step in achieving this is only to include operators that lead to different physical effects, i.e. to remove redundant operators.
A trivial example of this is to remove terms from the Lagrangian that are related by partial integration as they (given suitable boundary conditions) lead to the same action.

Less trivially, it is common in Effective Field Theories (EFTs) to employ the equations of motion (EOM) or field redefinitions to simplify and find a non-redundant operator base~\cite{Arzt:1993gz, Criado:2018sdb} (for some examples where this is done explicitly see, Higgs EFT~\cite{Feruglio:1992wf, Buchalla:2013rka}, Standard Model EFT~\cite{Buchmuller:1985jz,Grzadkowski:2003tf, Aguilar-Saavedra:2008nuh, Grzadkowski:2010es, Elias-Miro:2013gya,Jaeckel:2013uva}, or axion EFTs~\cite{Georgi:1986df,Weinberg:1996kr,Jaeckel:2013uva,Bauer:2016ydr,Bauer:2016zfj,GrillidiCortona:2015jxo,DiLuzio:2020wdo}).\footnote{Field redefinitions have also been used in the context of exact renormalization group equations to simplify the description and to elucidate the physical content, e.g.~\cite{Gies:2001nw,Gies:2002hq,Jaeckel:2002rm,Pawlowski:2005xe,Ihssen:2023nqd}.}
While the foundations for these procedures have been laid long ago in seminal papers~\cite{Chisholm:1961tha, Kamefuchi:1961sb, Divakaran:1963yxz, Kallosh:1972ap, Salam:1971sp, Ball:1993zy,Nambu:1968rr,Bergere:1975tr} 
we believe it useful to give a coherent and hopefully simple description of the non-trivial steps taken by doing so, using a language common in current phenomenological studies.

We find that many of the subtleties regarding the applications of field redefinitions can already be discussed at the classical level, on which we focus in this work.
For this purpose, we set up a theoretical basis based on classical objects instead of working with the semi-classical approximation of quantum ones.
This provides an alternative perspective in which concepts that might otherwise seem essential to the formalism (such as scattering amplitudes, asymptotic states, the infinite past/future limit, etc.) are unnecessary.
This is the first time that these results have been presented in this form, to the best of our knowledge.

Let us briefly outline the steps taken.
In Section~\ref{sec:lsz}, we present a classical analogue to the key result that allows the construction of redefinition-invariant observables in quantum field theory: the LSZ formula.
This section also sets the notation and essential concepts for the rest of the paper.
In Section~\ref{sec:interpolating}, we discuss two senses in which physics is invariant under redefinitions.
Firstly, all physical quantities must be independent of the parametrization used in field space.
However, one should be careful to redefine not only the action but also all observables in the theory.
Secondly, a stronger result is also possible: thanks to the LSZ theorem, there is a set of observables that is invariant under field redefinitions. Importantly, this includes cross-sections.
One can redefine the action but not these observables, and they will still take the same values.
In Section~\ref{sec:mechanical}, we use a mechanical toy model to illustrate the main points of the previous sections.
In Section~\ref{sec:non-invertible}, we study the issues that arise when field redefinitions are used beyond their range of applicability.
Typically, spurious solutions appear.
Some might be eliminated through additional boundary conditions, but not all can.
Some solutions to the original theory might also be lost.
In particular, we also consider the case of extended field solutions, such as solitons, where the elimination of spurious extra solutions is not always successful.
We conclude in Section~\ref{sec:conclusions} and briefly look towards the quantum theory.


\section{Classical version of the LSZ theorem}
\label{sec:lsz}

In quantum field theory, the key fact that ensures the invariance of scattering amplitudes under field redefinitions is the LSZ theorem. More precisely, the LSZ theorem links scattering to poles in correlation functions, and it is those that do not change under a field redefinition.
While intuition dictates that a classical version of LSZ should exist, such a result is not trivially obtained from the quantum case.
In fact, the LSZ formula deals with scattering amplitudes between asymptotic multi-particle states, which do not really exist classically. 
Thus, the most one can get by taking the classical limit on it is a tree-level semi-classical formula involving such states.
Instead, in this section, we construct and prove a version of this result that relies purely on classical concepts.

\subsection{Physical picture}
\label{sec:lsz-physical}

We consider a classical field theory for a field $\phi$ with action $S[\phi]$.
For simplicity, we assume that $\phi$ is a single scalar field, although all results here may be readily extended to collections of fields with non-vanishing spin.
We follow the usual setup employed in the study of quantum field theories and introduce an additional coupling of some local combination of the fields $F[\phi]$ to an external source $J$ so that the total action of the theory is given by
\begin{equation}
    S_J[\phi] = S[\phi] + F^x[\phi] J_x,
    \label{eq:S-FJ}
\end{equation}
where we have used DeWitt notation, in which the spacetime point is denoted as a regular index:
\begin{equation}
    \psi^x \equiv \psi_x \equiv \psi(x),
    \qquad 
    \phi^x \psi_x \equiv \int d^dx \, \phi(x) \psi(x).
\end{equation}
We will generally use this notation throughout this paper, but we will switch to the more traditional notation $\psi(x)$, whenever it helps clarify the meaning of a given formula.

The source $J$ is typically regarded as a formal parameter to obtain correlation functions through functional derivatives of a generating functional.
It can also be given a more physical meaning in two ways. Firstly, it can be viewed as an external field driving $\phi$ to set it up for the conditions of an experiment.
For example, setting $J$ to a combination of plane waves simulates a scattering process.
Secondly, it can be used to model a situation where the field can be measured only through the combination $F[\phi]$, but not directly.
For this purpose, one can promote $J$ to a dynamical field, playing the role of a measurement apparatus, and observe the effects of $\phi$ on it.

The classical dynamics of the $\phi$ field are governed by the following equation of motion
\begin{equation}
    \left(
        \frac{\delta S}{\delta \phi^x}
        + \frac{\delta F^y}{\delta \phi^x} J_y
    \right)_{\phi = \phi_J}
    = 0.
    \label{eq:J-eom}
\end{equation}
Clearly, the solution $\phi_J$ to this equation depends on the source $J$.
By focusing on the properties of $\phi_J$, we automatically incorporate the physical picture in which one can only act on the system through the source term $F^x J_x$.
One needs to restrict the set of available observables to introduce the idea that one can only measure the system through this term.
The only valid observables should be derived from the self-interactions of $J$, viewed as an external field, once $\phi$ has been integrated out. In other words, in the classical theory, it is evaluated at its solution to the equation of motion, $\phi_{J}$, in the presence of the source $J$.
The action containing those self-interactions is
\begin{equation}
    W[J] = S_J[\phi_J].
\end{equation}
All observables must be derived from $W[J]$.
We assume that both $\phi_J$ and $W[J]$ can be expanded as power series in $J$.
They are thus fully determined by their functional derivatives at $J = 0$,
\begin{align}
    \langle\phi(x) F(x_1) \cdots F(x_n) \rangle_S
    &\equiv
    \left. 
        \frac{\delta^n \phi_J(x)}{\delta J(x_1) \cdots \delta J(x_n)}
    \right|_{J = 0},
    \label{eq:phi-n-definition}
    \\
    \left<F(x_1) \cdots F(x_n) \right>_S
    &\equiv
    \left. 
        \frac{\delta^n W}{\delta J(x_1) \cdots \delta J(x_n)}
    \right|_{J = 0}.
    \label{eq:W-n-definition}
\end{align}
Here, the left-hand side is to be read as a definition where the notation 
is chosen to mimic the one used in quantum field theory (see Section~\ref{sec:quantum}, where we discuss the connection to quantum field theory in more detail). Moreover, we want to emphasise that these functions are computed using the action $S$ and the combination of fields $F$ in the source term.
We note that later $F$ will be our field redefinition's interpolating field (see next subsection).

The above notation also makes explicit that $\langle\phi(x)F(x_1)\cdots\rangle_S = \langle F(x) F(x_1) \cdots \rangle_S$ when $F[\phi] = \phi$.
This is the case because we can directly derive   $\phi_J = \delta W/\delta J$ by using the equations of motion.
We will call the $\left<\cdots\right>_S$ Green functions since they play the role of the connected Green functions in a quantum theory.
Intuitively, $\langle\phi(x) F(x_1) \cdots F(x_n)\rangle_S$ measures the response of the field $\phi$ at the point $x$ to the source $J$ acting at points $x_1$, \ldots, $x_n$,
while $\langle F(x_1) \cdots F(x_n)\rangle_S$ is the strength of the self interaction $J(x_1) \cdots J(x_n)$ induced by the field $\phi$.

Below, we will study the dependence of $\phi_J$ and $W[J]$ on the Fourier modes of $J$.
Specifically, we will see that, for a wide variety of possibilities for $F$, the positions and residues of the poles $\phi_J$ and $W[J]$ have in momentum space are independent of the concrete choice of $F$.
The situation is analogous to that of a driven anharmonic oscillator. When the driving force contains a mode with the resonant frequency---in our case, when the source has an on-shell mode, with $p^2 = m^2$, where $m$ is the mass of the field---the amplitude of the oscillations diverges. That is, both the solution to the EOM and the Green functions have a pole when the source goes on-shell.\footnote{It should be noted that this is not the only pole: additional ones may be present, at the points at which sums of frequencies contained in $J$ reach the frequency of resonance of the oscillator.}
This pole is reached for a general coupling to the source $JF[\phi]$, as long as $F[\phi]$ features an oscillation at the same frequency as $\phi$. 
All other frequencies in $F[\phi]$ are not resonant and do not change the pole.

\bigskip

For convenience, we will work with the Fourier transforms $\langle\tilde{\phi}(p) \tilde{F}(p_1) \cdots \rangle_S$ and $\langle\tilde{F}(p_1) \cdots \rangle_S$ of $\langle\phi(x) F(x_1) \cdots \rangle_S$ and $\langle F(x_1) \cdots \rangle_S$.
From them, one may reconstruct the full solution $\phi_J$ and the functional $W[J]$ through
\begin{align}
    \tilde{\phi}_J(p)
    &=
    \sum_{n = 0}^\infty \frac{1}{n!}
    \int d^4p_1 \cdots d^4p_n \;
    \langle\tilde{\phi}(p) \tilde{F}(p_1) \cdots \tilde{F}(p_n)\rangle_S \;
    \tilde{J}(p_1) \cdots \tilde{J}(p_n).
    \label{eq:phi-J-expansion}
    \\
    W[J]
    &=
    \sum_{n = 0}^\infty \frac{1}{n!}
    \int d^4p_1 \cdots d^4p_n \;
    \langle\tilde{F}(p_1) \cdots \tilde{F}(p_n)\rangle_S \;
    \tilde{J}(p_1) \cdots \tilde{J}(p_n).
    \label{eq:W-J-expansion}
\end{align}
where $\tilde{\phi}_J$ and $\tilde{J}$ are the Fourier transforms of $\phi_J$ and $J$, respectively.

\subsection{Interpolating fields}
\label{sec:lsz-interpolating}

For the results of this section to be applicable, we need to impose
some general conditions 
on the functional $F$. 
We follow the nomenclature employed in quantum field theory and call such a suitable functional an interpolating field (see e.g.~\cite{Bain:2000lua, Collins:2019ozc}).
In the quantum case, the defining condition for interpolating fields is that the overlap $\left<0\right| F[\phi] \left|p\right>$ of the states it creates with one-particle states is non-vanishing.
Here, we provide a definition purely in classical terms.
The functional derivative of $F$ with respect to $\phi$ may be viewed as a linear differential operator acting on test functions $\psi$.
Our requirement is that this operator is non-singular for all $\phi$.

\begin{definition}\label{def:interpolating-field}
A local functional of the fields $F[\phi]$ is called an \textbf{interpolating field} whenever
\begin{equation}
    \frac{\delta F^y}{\delta \phi^x} \psi_y = 0
    \quad \iff \quad \psi(x) \equiv 0.
    \label{eq:F-nonsingular}
\end{equation}
\end{definition}

The fact that $\delta F/\delta \phi$ is non-singular in the sense of Eq.~\eqref{eq:F-nonsingular} is equivalent to the existence of an inverse differential operator $(\delta F / \delta \phi)^{-1}$.
In fact, using the inverse function theorem, one can see that $F$ is a one-to-one mapping, and an inverse functional $F^{-1}[\phi]$ should exist.
These properties will become important in Section~\ref{sec:interpolating}, where we will require that field redefinitions are of the form $\phi = F[\varphi]$, with $F$ being an interpolating field.

Beyond Eq.~\eqref{eq:F-nonsingular}, we require that $F$ is local.
This means that $F^x[\phi]$ may only depend on a \emph{finite} number of derivatives $\partial^N\phi(x)$ of the field at the point $x$. 
Whenever derivatives are present in $F$, the left-hand side of Eq.~\eqref{eq:F-nonsingular} becomes a differential equation that may have non-trivial solutions, which forbids it from being an interpolating field.
However, if we are working with the field theory only through perturbation theory, Eq.~\eqref{eq:F-nonsingular} only has to be true perturbatively, allowing for interpolating fields with derivatives.
Thus, there are two main ways to construct an interpolating field:
\begin{enumerate}
    \item $F$ is ultralocal, meaning it has no derivatives so that it can be written as $F^x[\phi] = f(\phi(x))$.
    Additionally, $f$ is required to be invertible function ($f'(\phi) \neq 0$ everywhere).
    An example of this is $F^x[\phi] = \phi(x) + g \phi^3(x)$.

    \item $F$ is a power series in a perturbative parameter $\lambda$, with the $\lambda^0$ order being ultralocal and invertible. Only at order $\lambda$ and higher may contain a finite number of derivatives.
    An example of this is $F^x[\phi] = \phi(x) + \lambda \, \square \phi(x)$.
    It should be noted that, in order for any result relying on the interpolating nature of $F$ to work, all quantities, including the action and all observables, should be perturbative series in $\lambda$, too.
\end{enumerate}
The second situation is commonly encountered when working with effective field theories, where $\lambda$ is the perturbative parameter of the effective theory, usually proportional to the inverse of the cutoff scale.
In the literature, field redefinitions are usually assumed to be either of the first type~\cite{Chisholm:1961tha,Kamefuchi:1961sb,Divakaran:1963yxz,Kallosh:1972ap,Salam:1971sp,Nambu:1968rr}
or the second one~\cite{Bergere:1975tr,Ball:1993zy,Arzt:1993gz,Criado:2018sdb}.
Here, we discuss both possibilities in a unified way through the general definition~\ref{def:interpolating-field}.
We also note that a typical assumption in the quantum formulations is that $F$ is a polynomial, or a least a power series, in the fields.
This is not required in the results we present here.

\subsection{Results}
\label{sec:lsz-results}

With these preliminaries, we can present several results for a generic classical field theory in the same spirit as the LSZ formula for a quantum one.
We will prove them in Appendix~\ref{sec:proofs}.

We make a common set of assumptions for all results:
\begin{enumerate}
    \item The action $S$ is local.

    \item The trivial field configuration $\phi \equiv 0$ is a solution to the equation of motion~\eqref{eq:J-eom} with $J=0$.
  We assume this is the leading order in the expansion of $\phi_J$ in power of $J$.
  That is, $\langle\phi(x)\rangle_S = 0$. This choice usually corresponds to the lowest energy state, i.e. for $J = 0$ we are in the QFT vacuum state.

    \item $F$ is an interpolating field.

    \item The linearized equation of motion arising from $S$ is the Klein-Gordon equation. Explicitly:
        \begin{equation}
          \left.
            \frac{\delta^2 S}{\delta\tilde{\phi}^p \delta\tilde{\phi}^q}
          \right|_{\phi = 0}
          =
          \delta^d(p + q)
          (p^2 - m^2)
          (1 - \lambda E(p)),
          \label{eq:classical-lsz-assumptions-free}
        \end{equation}
        where $E(p)$ is a polynomial in $p$ that may only be non-zero if $F$ is perturbative with small parameter $\lambda$.
\end{enumerate}
Our first result is about the on-shell poles of the $\langle\phi(p) F(p_1) \cdots\rangle_S$ functions.

\begin{theorem}
    \label{th:lsz-1}
    In the limit $p_1^2 \to m^2$, $\cdots$, $p_n^2 \to m^2$, we have the asymptotic relation
    \begin{equation}
      \langle\tilde{\phi}(p_1) \tilde{F}(p_2) \cdots \rangle_S
      \sim
      \frac{\sqrt{Z_\phi(p_1)}}{p_1^2 - m^2}
      \frac{\sqrt{Z_F(p_2)}}{p_2^2 - m^2}
      \cdots
      \frac{\sqrt{Z_F(p_n)}}{p_n^2 - m^2}
      \,
      \delta^4(p_1 + \cdots + p_n)
      \,
      \mathcal{A}(p_1, \cdots, p_n),
      \label{eq:th-lsz-1}
    \end{equation}
    with the following properties:
    \begin{itemize}
        \item $\sqrt{Z_F(p)} \neq 0$ is a polynomial in $p$ which, in general, depends both on $F$ and $S$.
        \item $\mathcal{A}(p_1, \cdots, p_n)$ is an analytic function of the momenta, except possibly when sums of subsets of them go on-shell $(p_{i_1} + p_{i_2} + \cdots)^2 \to m^2$.
        It depends on $S$ but not on $F$.
        Additionally,
        \begin{equation}
            \mathcal{A}(p_1, p_2) = p_1^2 - m^2.
        \end{equation}
    \end{itemize}
\end{theorem}

One may re-formulate this theorem in intuitive terms as follows: if the source is taken to be a combination of plane waves with momenta $p_1$, $\ldots$, $p_n$, then $\langle\phi(p_1) F(p_2) \cdots F(p_n)\rangle_S$ has a pole when all of these momenta are put on-shell.
Near the pole, the field splits into factors: a normalization $\sim Z^{n/2}$ related to the sourced field $F$, and an amplitude $\mathcal{A}$ completely determined by the action only.

\begin{theorem}
    \label{th:lsz-2}
    In the limit $p_1^2 \to m^2$, $\cdots$, $p_n^2 \to m^2$, we have the asymptotic relation
    \begin{equation}
      \langle\tilde{F}(p_1) \cdots \tilde{F}(p_n)\rangle
      \sim
      \frac{\sqrt{Z_F(p_1)}}{p_1^2 - m^2}
      \cdots
      \frac{\sqrt{Z_F(p_n)}}{p_n^2 - m^2}
      \,
      \delta^4(p_1 + \cdots + p_n)
      \,
      \mathcal{A}(p_1, \cdots, p_n).
      \label{eq:th-lsz-2}
    \end{equation}
\end{theorem}

\bigskip

Theorems~\ref{th:lsz-1} and \ref{th:lsz-2} ensure that, up to factors of $\sqrt{Z_F}$, the positions and residues of the on-shell poles of the field solution and Green functions are independent of the interpolating field used to act on the system and perform measurements.

\subsection{Connection to quantum field theory}
\label{sec:quantum}

The definitions and results presented above, which, in principle, apply to classical field theories, can be directly related to similar ones in quantum field theories at the tree level.
Although the tree-level approximation is widely known to be related to the classical limit of quantum theories, this is usually a semi-classical statement~\cite{Lee:1967ug,Boulware:1968zz,Nambu:1968rr,Brown:1992ay}, where this means that this approximation is applied to objects that can only be defined in quantum field theory, such as quantum states and observables, viewed as elements and operators in a Hilbert space, respectively.
To build a bridge to QFT, we here provide a connection between some of these objects and those we have defined above, which instead refer to properties of the classical fields $\phi$ and $J$.

The tree-level approximation to the generating functional in a quantum theory is given by\footnote{In other words, the only path taken is the one that corresponds to a saddle point of the equation and, therefore, fulfils the classical equation of motion.}
\begin{equation}
    Z[J] = \frac{1}{{\mathcal{N}}}\int \mathcal{D}\phi \, e^{i S_J[\phi]}
    = e^{i S_J[\phi_J]} + O(\hbar),
\end{equation}
where we have included a normalization factor ${\mathcal{N}}$ for convenience.
The functional $W[J] = S_J[\phi_J]$ generates the functional for tree-level connected Green functions.
Therefore, the relation between the (connected) quantum (left side) and classical (right side) Green functions is
\begin{equation}
    \langle 0 | 
    T\{F^{x_1}[\Phi] \cdots F^{x_n}[\Phi]\}
    | 0 \rangle_c
    =
    i^n \langle F(x_1) \cdots F(x_n) \rangle_S
    + O(\hbar).
\end{equation}
More precisely, the left-hand side denotes the connected part of the vacuum expectation value for the time-ordered product of the operators $F^{x_i}[\Phi]$, with $\Phi(x)$ being the operator representing the quantum field $\phi$.
Similarly,
\begin{equation}
    \langle 0 | 
    T\{\Phi^x  F^{x_1}[\Phi] \cdots F^{x_n}[\Phi]\}
    | 0 \rangle_c
    =
    i^n \langle \phi(x) F(x_1) \cdots F(x_n) \rangle_S
    + O(\hbar).
\end{equation}
Using this relation in theorems~\ref{th:lsz-1} and~\ref{th:lsz-2}, one obtains a statement about the momentum dependence of these time-ordered vacuum expectation values in the on-shell limit.
The quantum LSZ formula gives rise to the same dependence in this limit.
Our results provide an alternative proof of this structure at tree level.

Additionally, by comparing Eqs.~\eqref{eq:th-lsz-1} and~\eqref{eq:th-lsz-2} to their quantum version, we can extract the quantum objects that correspond to the functions $Z_F(p)$ and $\mathcal{A}(p_1, \cdots, p_n)$.
On the one hand theorem~\ref{th:lsz-2} gives,
\begin{equation}
    \langle \tilde{F}(p) \tilde{F}(q) \rangle_S
    \sim \frac{\delta^d(p + q)}{p^2 - m^2} Z_F(p).
\end{equation}
On the other hand, the K\"all\'en-Lehmann representation specifies
\begin{equation}
    \langle 0 | T \{ \tilde{F}[\Phi](p) \tilde{F}[\Phi](q) \}
    | 0 \rangle
    \sim \frac{\delta^d(p + q)}{p^2 - m^2}
    \left| \langle 0 | F[\Phi](x_0) | \mathbf{p} \rangle \right|^2,
\end{equation}
where we have denoted $p = (E, \mathbf{p})$ and $x_0$ is an arbitrary spacetime point.
Comparing both equations, we obtain 
\begin{equation}
    |\langle 0 | F[\Phi](x_0) | \mathbf{p} \rangle|
    = \sqrt{Z_F(p)} + O(\hbar).
\end{equation}
The fact that $Z_F(p) \neq 0$, which comes from the assumption that $F$ is an interpolating field in the classical sense, as shown in the Appendix~\ref{sec:proofs}, implies that $F[\Phi]$ is an interpolating field in the quantum sense. 
That is, $\langle 0 | F[\Phi](x_0) | \mathbf{p} \rangle \neq 0$.
For the $\mathcal{A}$ functions, we write $p_i = (E_i, \mathbf{p}_i)$, and assume that $E_1$, $\cdots$, $E_k > 0$ and $E_{k+1}$, $\cdots$, $E_n < 0$
Then, we have
\begin{equation}
    \langle \text{in}; \mathbf{p}_1, \cdots, \mathbf{p}_k |
    -\mathbf{p}_{k+1}, \cdots, -\mathbf{p}_n; \text{out}\rangle
    =
    \mathcal{A}(p_1, \cdots, p_n)
    + O(\hbar),
\end{equation}
where $|\cdots; \text{in}\rangle$ and $|\cdots; \text{out}\rangle$ denote the asymptotic multi-particle in and out states.
In physical terms, $Z_F$ quantifies the strength of the oscillation in $F[\phi]$ with the same resonant frequency as $\phi$.

\section{Field redefinitions with interpolating fields}
\label{sec:interpolating}

In this section, we consider a field redefinition of the form
\begin{equation}
    \phi = F[\varphi],
\end{equation}
where $F$ is an interpolating field.
After the redefinition, one ends up with a theory whose action $\hat{S}$ is given by
\begin{equation}
    \hat{S}[\varphi] = S[F[\varphi]].
\end{equation}
Naively, the theories defined by $S$ and $\hat{S}$ should be equivalent.
However, some care is needed regarding the observables under consideration.
One has to either redefine all observables together with the action or consider a subset of all observables that remain invariant even without this additional redefinition, thanks to the classical LSZ theorem proven above.

\subsection{Redefining observables}
\label{sec:redefining-observables}

Consider a theory defined by the action $S$ with a trivial source term $\phi^x J_x$ and the corresponding action with a source after the redefinition:
\begin{equation}
    S_J[\phi] = S[\phi] + \phi^x J_x,
    \qquad \qquad
    \hat{S}_J[\varphi] \equiv S_J[F[\varphi]] = \hat{S}[\varphi] + F^x[\varphi] J_x.
\end{equation}
We are interested in the solutions $\phi_J$ and $\varphi_J$ to the EOMs derived from these two actions.
We will see that these solutions are in one-to-one correspondence.
The EOMs are given by
\begin{equation}
    0 = \left.\frac{\delta S_J}{\delta \phi^x}\right|_{\phi_J},
    \qquad \qquad
    0 = \left.\frac{\delta \hat{S}_J}{\delta \varphi^x}\right|_{\varphi_J}.
\end{equation}
We may use the relation between $S$ and $\hat{S}$ and the interpolating property of $F$ to re-write these equations in a different form:
\begin{equation}
    0 = 
    \left.\frac{\delta G^y}{\delta \phi^x}\right|_{\phi_J}
    \left.\frac{\delta \hat{S}_J}{\delta \varphi^y}\right|_{G[\phi_J]},
    \qquad \qquad
    0 = 
    \left.\frac{\delta F^y}{\delta \varphi^x}\right|_{\varphi_J}
    \left.\frac{\delta S_J}{\delta \phi^x}\right|_{F[\varphi_J]},
\end{equation}
where we have denoted $G \equiv F^{-1}$ for simplicity.
Noting that $F$ and $G$ are interpolating fields again, we know that the two differential operators $\delta G/\delta\phi$ and $\delta F/\delta\varphi$ are non-singular, so the two equations are equivalent to
\begin{equation}
    0 = 
    \left.\frac{\delta \hat{S}_J}{\delta \varphi^y}\right|_{G[\phi_J]},
    \qquad \qquad
    0 = 
    \left.\frac{\delta S_J}{\delta \phi^x}\right|_{F[\varphi_J]}.
\end{equation}
In this form, it is clear that any solution $\varphi_J$ of the new EOM induces a solution of the old EOM through $\phi_J = F[\varphi_J]$, and vice versa, any solution $\phi_J$ to the new EOM induces a solution to the old one through $\varphi_J = G[\phi_J] = F^{-1}[\phi_J]$.
In this sense, the dynamics are unchanged.

However, there is one point where we need to be careful in a slightly trivial way.
In general, $\phi^x\neq \varphi^x$.
Considering the same observable, we need to use the relation between the two fields, i.e. $\varphi^x = G^x[\phi]$.
This holds for any observable in the theory.
Let $\mathcal{O}[\varphi]$ denote a generic observable.
In general, we will have $\mathcal{O}[\varphi_J] \neq \mathcal{O}[\phi_J]$.
Instead, the correct relation is
\begin{equation}
    \mathcal{O}[\phi_J] = \hat{\mathcal{O}}[\varphi_J],
    \qquad
    \text{where } \hat{\mathcal{O}}[\varphi] \equiv \mathcal{O}[F[\varphi]].
    \label{eq:observable-redefinition}
\end{equation}

This is automatically obtained if one only considers the valid observables defined in Section~\ref{sec:lsz}: the Green functions derived from $W[J]$.
Let us differentiate between the $W[J]$ for the original and the redefined theory by denoting it $W[J]$ and $\hat{W}[J]$, respectively.
We then have
\begin{equation}
    W[J] = S_J[\phi_J] = S_J[F[\varphi_J]] = \hat{S}_J[\varphi_J] = \hat{W}[J],
\end{equation}
where we have used that $\phi_J = F[\varphi_J]$.
This leads to the following equivalence of Green functions:
\begin{equation}
    \langle \phi(x_1) \cdots \phi(x_n)\rangle_S
    =
    \langle F(x_1) \cdots F(x_n)\rangle_{\hat{S}}.
\end{equation}
This equivalence only holds because we have kept track of the redefinition in the source terms.
If, instead of doing that, we had naively set the source term in the redefined theory to $\varphi^x J_x$, as
\begin{equation}
    \bar{S}_J[\varphi] = \hat{S}[\varphi] + \varphi^x J_x, 
\end{equation}
we would have obtained a different set of Green functions.
Using $\bar{W}[J]$ to denote the corresponding functional for the theory defined by $\bar{S}_J$, we would, in general, have 
\begin{equation}
    W[J] \neq \bar{W}[J],
    \qquad \qquad
    \langle \phi(x_1) \cdots \phi(x_n)\rangle_S
    \neq
    \langle \varphi(x_1) \cdots \varphi(x_n)\rangle_{\hat{S}}.
    \label{eq:neq-Green-functions}
\end{equation}

\subsection{Redefinition-invariant observables}

So far, we have only provided a more precise formulation that physical quantities are independent of how the field space is parametrized.
A stronger statement can be obtained using the classical LSZ theorems in Section~\ref{sec:lsz}.
We wish to turn the inequality Eq.~\eqref{eq:neq-Green-functions} into equality through suitable modifications.
The correct way to do this is to replace Green functions with the classical LSZ amplitudes $\mathcal{A}$.
We have 3 sets of normalization factors $Z$ and amplitudes $\mathcal{A}$ arising from the three different theories we have considered in Section~\ref{sec:redefining-observables}:
\begin{itemize}
    \item In the original theory, we have $Z_\phi$ and $\mathcal{A}$, arising from the on-shell limit of the Green functions generated by $W[J]$.
    \item In the redefined theory with a redefined source term, we have $\hat{Z}_F$ and $\hat{\mathcal{A}}$, arising from the Green functions of $\hat{W}[J]$.
    \item In the theory with redefined action but a naive source term, we have $\bar{Z}_\varphi$, $\bar{\mathcal{A}}$, coming from the Green functions of $\bar{W}[J]$.
\end{itemize}

The amplitudes constitute an observable that remains invariant even without applying the redefinition to it.
In other words, if one is only interested in them, one can perform the naive redefinition procedure, applying it only to the action and forgetting about its effects on the source term.
The proof is simple.
Due to the identity $W[J] = \hat{W}[J]$, which makes all Green functions the same in both theories, we have that $Z_\phi(p) = \hat{Z}_F(p)$ and $\mathcal{A}(p_1, \cdots) = \hat{\mathcal{A}}(p_1, \cdots)$.
Now, the only difference between the $\hat{W}[J]$ and $\bar{W}[J]$ theories is the source terms.
But, according to theorem~\ref{th:lsz-1}, the amplitudes only depend on the action, and so $\hat{\mathcal{A}}(p_1, \cdots) = \bar{\mathcal{A}}(p_1, \cdots)$.
Therefore,
\begin{equation}
    \mathcal{A}(p_1, \cdots, p_n)
    =
    \bar{\mathcal{A}}(p_1, \cdots, p_n).
\end{equation}
One may reformulate this result in a form closer to that of Eq.~\eqref{eq:neq-Green-functions} as
\begin{equation}
    \frac{\langle\tilde{\phi}(p_1) \cdots \tilde{\phi}(p_n)\rangle_S} {\sqrt{Z_\phi(p_1) \cdots Z_\phi(p_n)}}
    \sim
    \frac{\langle\tilde{\varphi}(p_1) \cdots \tilde{\varphi}(p_n)\rangle_{\hat{S}}}{\sqrt{\bar{Z}_\varphi(p_1) \cdots \bar{Z}_\varphi(p_n)}}, \quad{\rm for} \quad p_1^2, \cdots p_n^2 \to m^2.
    \label{eq:on-shell-equivalence}
\end{equation}

A natural recipe for working with redefinition-invariant quantities in classical field theories arises from this result.
First, one should only measure the system's response to sources that are linear combinations of (perhaps near) on-shell plane waves. In other words, the equivalence between theories with and without redefined source terms holds for cross sections and related quantities.\footnote{In general, correlation functions differ when not LSZ amputated, see the example in Eq.~\eqref{eq:example-3-point}. It is not a priori clear to us that those are not measurable.}
Second, the two-point Green function can be used to obtain the normalization factor $Z$.
This factor should be employed to normalize the on-shell Green functions as in Eq.~\eqref{eq:on-shell-equivalence}.
Then, one can perform field redefinitions in the action without additional steps, and all the normalized measurements will remain invariant.

\subsection{Using equations of motion vs field redefinitions}\label{sec:eom}

Finally let us comment on another commonly employed technique to remove redundant operators from the Lagrangian (see~\cite{Buchmuller:1985jz,Feruglio:1992wf,Grzadkowski:2003tf,Aguilar-Saavedra:2008nuh,Grzadkowski:2010es,Elias-Miro:2013gya,Buchalla:2013rka,Jaeckel:2013uva,Bauer:2016ydr,Bauer:2016zfj} for some examples where this is done): using the equation of motion on suitable terms in the Lagrangian.
This can, at best, be an approximation.
A simple counterexample to this being an exact procedure is the free Dirac field or, for that matter, any free field theory. One can easily convince oneself that inserting the solution to the equation of motion into the Lagrangian yields a vanishing result and, therefore, no useful Lagrangian.

It is, therefore, worthwhile to understand how one can use this as an approximate procedure and the corresponding limitations. This is, of course, not new. It has been discussed (in the context of quantum field theory) in Ref.~\cite{Arzt:1993gz} and more recently emphasized in Ref.~\cite{Criado:2018sdb}. Following our spirit of collecting such results into a simple and coherent picture, let us recall the main argument.

The point is that, at leading order, an effective operator can be removed by a variable change, i.e. a field redefinition.
To see this we follow~\cite{Arzt:1993gz,Criado:2018sdb} and consider a perturbative redefinition
\begin{equation}
    \phi=F[\phip]=\phip+\lambda H[\phip]
\end{equation}
where $\lambda$ is a small parameter, all quantities in the theory are regarded as power series, i.e. we are considering an interpolating field.
This can now be inserted into the action and expanded in the perturbative parameter,
\begin{equation}
   \hat{S}[\phip]= S[F(\phip)] = S[\phip] + \lambda H[\phip] \frac{\delta S}{\delta \phi}\bigg|_{\phi=\phip} + \lambda^2 H[\phip]^2\frac{\delta^{2}S[\phi]}{(\delta \phi)^{2}}\bigg|_{\phi=\phip}+\ldots\,.
\end{equation}
The first term on the right-hand side is the original action as a function of the new field $\phip$. The second is proportional to the equation of motion. It is, therefore, clear that at the leading order, we can now remove any operator that multiplies the equation of motion by choosing a suitable $\lambda H$.
Evidently, this procedure only works at leading order. 
The second (and higher) terms provide corrections.

\section{Mechanical examples}
\label{sec:mechanical}

\subsection{Perturbative redefinition in the harmonic oscillator}
\label{sec:harmonic-oscillator}

To illustrate the issues discussed so far, we consider a toy mechanical example (that is, a field theory in 1+0 dimensions) in which the only degree of freedom is the 1-dimensional position of a single particle.
To match the usual notation, we employ the symbols $x$ and $y$ to denote the position before and after the redefinition.
They play the role of the fields $\phi$ and $\varphi$ above.
Similarly, the time $t$ corresponds to the spacetime point $x$ in the previous discussion.
For simplicity, we focus first on the theory without a source.
The action and the corresponding EOM are
\begin{equation}
    S[x] = \frac{1}{2} \int dt \, (\dot{x}^2 + \omega^2 x^2),
    \qquad
    \qquad
    \ddot{x} + \omega^2 x = 0.
    \label{eq:mechanical-original-eom}
\end{equation}
The redefinition we consider is
\begin{equation}
    x = F[y] = y + \lambda \dot{y}.
\end{equation}

For $F$ to be an interpolating field, one needs to interpret $\lambda$ as a perturbative parameter and expand all quantities as a power series in it.
This is the second option in Section~\ref{sec:lsz} to construct an interpolating field.
In Section~\ref{sec:non-invertible}, we will return to this example and discuss the issues arising from a non-perturbative treatment.
Sticking to perturbation theory for now, we may explicitly construct the inverse of $F$:
\begin{equation}
    y = G[x] \equiv F^{-1}[x]
    = x - \lambda \dot{x} + \lambda^2 \ddot{x} - \cdots
    = \sum_{k = 0}^\infty (-\lambda)^k x^{(k)}.
\end{equation}
We can also compute the functional derivative $\delta F/\delta y$ at this point.
As we know, this is a non-singular linear differential operator.
The action of both this operator and its inverse on a test function $f(t)$ can also be explicitly written as
\begin{equation}
    \frac{\delta F^\tau}{\delta y^t} f_\tau
    = f_t - \lambda \dot{f}_t,
    \qquad \qquad
    \frac{\delta G^\tau}{\delta y^t} f_\tau
    = f_t + \lambda \dot{f}_t + \lambda^2 \ddot{f}_t + \ldots = \sum_{k = 0}^\infty \lambda^k f^{(k)}_t.
\end{equation}

After the redefinition, the action becomes
\begin{align}
    \hat{S}[y] = S[F(y)]
    &= \frac{1}{2} \int dt \left[
        \dot{y}^2 + \omega^2 y^2 
        + 2 \lambda (\dot{y} \ddot{y} + \omega^2 y \dot{y}) 
        + \lambda^2 (\ddot{y}^2 + \omega^2 \dot{y}^2)
    \right]
    \nonumber \\
    &= \frac{1}{2} \int dt \Big[
        \dot{y}^2 + \omega^2 y^2 
        + \lambda^2 (\ddot{y}^2 + \omega^2 \dot{y}^2)
    \Big]
    + \lambda \Big[\dot{y}^2 + \omega^2 y^2\Big]_{\text{boundary}}
\end{align}
The corresponding EOM is
\begin{equation}
    (1 - \lambda^2 \partial_t^2) (\ddot{y} + \omega^2 y) = 0.
    \label{eq:mechanical-redefined-eom}
\end{equation}
There is a perturbative one-to-one correspondence between the solutions
of the original EOM~\eqref{eq:mechanical-original-eom} and the redefined EOM~\eqref{eq:mechanical-redefined-eom}.
This can be checked by perturbatively solving the two equations and noting that the solutions match.
\begin{itemize}
    \item If $x$ solves the harmonic oscillator, so do its derivatives. This implies $y = F^{-1}(x)$ is also a solution to the harmonic oscillator, and thus to $\delta S'/\delta y = 0$. 
    \item If $y = y_0 + \lambda y_1 + \cdots$ solves $\delta \hat{S}/\delta y = 0$, then $y_0$ and $y_1$ solve the harmonic oscillator equation (because the corrections in $\delta \hat{S}/\delta y$ start at $\lambda^2$). But then, at every order $n$, we will have the harmonic oscillator equation for $y_n$ plus a second derivative of the harmonic oscillator equation for $y_{n-2}$. So all orders solve it, and then $x = F(y)$ does it too. The only way to avoid this is by having non-perturbative terms, e.g.  $e^{t/\lambda}$, in the solutions. Therefore, any extra solutions are always non-perturbative. 
\end{itemize}

Let us now include the source terms. Let $x_J$ and $y_J$ be the solutions to the equations of motion with sources:
\begin{equation}
    0 = \left.\left(
        \frac{\delta S}{\delta x^t} - J_t
    \right)\right|_{x = x_J},
    \qquad \qquad
    0 = \left.\left(
        \frac{\delta \hat{S}}{\delta y^t} - J_t + \lambda \dot{J}_t
    \right)\right|_{y = y_J}.
\end{equation}
By the same argument as before, there should be a one-to-one correspondence between them at the perturbative level, given by $x_J = F[y_J]$ and $y_J = F^{-1}[x_J]$. This can be seen explicitly. $x_J$ is easy to compute:
\begin{equation}
    x_J(t) = \int dE \frac{e^{-i E t}}{E^2 - \omega^2} \tilde{J}(E)
\end{equation}
For $y_J$, we can use a diagrammatic approach. In momentum space, there are $1/(E^2 - \omega^2)$ propagators, two-legged insertions of $-\lambda^2 E^2 (E^2 - \omega^2)$ and sources $(1 + i \lambda E) \tilde{J}(E)$, where $\tilde{J}(E) = \int dt \, e^{-i E t} J(t)$. Each two-legged insertion adds an additional propagator, so the net effect is just a $-\lambda^2 E^2$ factor. We thus have:
\begin{equation}
    y_J(t) = \int dE \frac{e^{-i E t}}{E^2 - \omega^2} \left[
        1 - \lambda^2 E^2 + \lambda^4 E^4 - \cdots
    \right] (1 + i \lambda E) \tilde{J}(E).
\end{equation}
Transforming back to $x$ space gives the solution to the original theory:
\begin{align}
    F_t[y_J] &= y_J(t) + \lambda \dot{y}_J(t)
    \nonumber \\
    &= \int dE \, (1 - i\lambda E) \frac{e^{-i E t}}{E^2 - \omega^2} \left[
        1 - \lambda^2 E^2 + \lambda^4 E^4 - \cdots
    \right] (1 + i \lambda E) \tilde{J}(E)
    \nonumber \\
    &= \int dE \frac{e^{-i E t}}{E^2 - \omega^2} \tilde{J}(E)
    = x_J(t).
\end{align}
Going in the other direction, we obtain:
\begin{align}
    F^{-1}_t[x_J]
    &= x_J(t) - \lambda \dot{x}_J(t) + \lambda^2 \ddot{x}_J(t) + \cdots
    \nonumber \\
    &= \int dE \left[
        1 + i \lambda E - \lambda^2 E^2+  \cdots
    \right] \frac{e^{-i E t}}{E^2 - \omega^2} \tilde{J}(E)
    = y_J(t).
\end{align}
We may also check that both solutions satisfy theorem~\ref{th:lsz-1}.
The $n$-point functions for $n > 2$ are all zero, so the only non-trivial case is
\begin{align}
    \langle \tilde{x}(E_1) \tilde{x}(E_2) \rangle 
    &= \frac{\delta(E_1 + E_2)}{E^2 - m^2},
    \\
    \langle \tilde{y}(E_1) \tilde{F}(E_2) \rangle 
    &= \frac{\delta(E_1 + E_2)}{E^2 - m^2}
    (1 - \lambda^2 E_2^2 + \lambda^4 E_2^4 + \cdots) (1 + i\lambda E_2).
\end{align}
Clearly, both functions satisfy the theorem since they have a pole at $E_2^2 \to m^2$.
The normalization factor for them is, however, different:
\begin{equation}
    \sqrt{Z_x(E)} = 1,
    \qquad \qquad
    \sqrt{\hat{Z}_F(E)}
    = (1 - \lambda^2 \omega_2^2 + \lambda^4 \omega_2^4 + \cdots) (1 + i\lambda E_2) 
\end{equation}

We can also consider the theory after the naive redefinition procedure outlined above.
If we perform the redefinition in the action but not in the source terms, the solution $\bar{y}_J$ is different from $y_J$:
\begin{equation}
    0
    =
    \left.\left(
        \frac{\delta S'}{\delta y^t} - J_t
    \right)\right|_{y = \bar{y}_J},
    \qquad
    \bar{y}_J(t) = \int dE \frac{e^{-i E t}}{E^2 - \omega^2} \left[
        1 - \lambda^2 E^2 + \lambda^4 E^4 - \cdots
    \right] \tilde{J}(E).
\end{equation}
However, in agreement with theorem~\ref{th:lsz-1}, it still has a pole at $E^2 \to \omega^2$.
The difference is again the normalization factor, which now reads:
\begin{equation}
    \sqrt{\hat{Z}_y(E)} = 1 + \lambda^2 \omega^2 + \lambda^4 \omega^4 + \cdots
\end{equation}

\subsection{Classical LSZ in the anharmonic oscillator}

To get a non-trivial application of the LSZ theorems with non-vanishing higher-point functions, we consider now two theories with the same action
\begin{equation}
    S[x] = \frac{1}{2} \int dt
    \left(\dot{x}^2 + \omega^2 x^2 + \frac{g}{3} x^3\right).
\end{equation}
But different source terms:
\begin{equation}
    S_J[x] = S[x] + x^t J_t
    \qquad \qquad
    \bar{S}_J[x] = S[x] + F^t[x] J_t,
\end{equation}
where $F^t[x] = x(t) + \lambda (\dot{x}(t) + x(t)^2/2)$.
As discussed before, we assume that $\lambda$ is a perturbative parameter to make $F$ an interpolating field.
The Green functions of these theories can be computed diagrammatically.
The Feynman rules for internal lines and vertices are
\begin{equation}
    \begin{tikzpicture}[baseline=-2pt]
        \draw (0, 0) -- (1, 0);
    \end{tikzpicture}
    =
    \frac{1}{E^2 - \omega^2},
    \qquad \qquad
    \begin{tikzpicture}[baseline=-2pt]
        \draw (0, 0) -- (-0.5, 0);
        \draw (0, 0) -- (0.25, 0.44);
        \draw (0, 0) -- (0.25, -0.44);
    \end{tikzpicture}
    =
    g.
\end{equation}
The rules for external vertices of both theories are different, corresponding to the two different interpolating fields $x$ and $F[x]$ coupled to the source in them.
We use the following notation for them:
\begin{equation}
    \text{for } S_J: \quad
    \begin{tikzpicture}[baseline=-2pt]
        \draw (0, 0) -- (1, 0);
        \draw[fill=black] (1, 0) circle (2pt);
    \end{tikzpicture}
    =
    1,
    \qquad \qquad \qquad
    \text{for } \bar{S}_J: \quad
    \begin{tikzpicture}[baseline=-2pt]
        \draw (0, 0) -- (1, 0);
        \draw[fill=white] (1, 0) circle (2pt);
    \end{tikzpicture}
    =
    1 + i \lambda E,
    \qquad
    \begin{tikzpicture}[baseline=-2pt]
        \draw (0, 0) -- (-0.4, 0.4);
        \draw (0, 0) -- (-0.4, -0.4);
        \draw[fill=white] (0, 0) circle (2pt);
    \end{tikzpicture}
    =
    \lambda.
\end{equation}

First, we compute the 2-point functions:
\begin{align}
    \langle \tilde{x}(E_1) \, \tilde{x}(E_2) \rangle_S &=
    \;
    \begin{tikzpicture}[baseline=-2pt]
        \draw (0, 0) -- (1, 0);
        \draw[fill=black] (0, 0) circle (2pt);
        \draw[fill=black] (1, 0) circle (2pt);
    \end{tikzpicture}
    \;
    = 
    \frac{1}{E_1^2 - \omega^2} \delta(E_1 + E_2)
    \\
    \langle \tilde{F}(E_1) \, \tilde{F}(E_2) \rangle_S &=
    \;
    \begin{tikzpicture}[baseline=-2pt]
        \draw (0, 0) -- (1, 0);
        \draw[fill=white] (0, 0) circle (2pt);
        \draw[fill=white] (1, 0) circle (2pt);
    \end{tikzpicture}
    \;
    = 
    \frac{(1 + i\lambda E_1)(1 + i\lambda E_2)}{E_1^2 - \omega^2} \delta(E_1 + E_2).
\end{align}
As expected from theorem~\ref{th:lsz-1}, they both have a pole (with a positive sign) at $E^2 \to \omega^2$, so we know that the physical frequency in both cases is $\omega$. The residue at this pole is different. The fact that $\bar{x}_J$ comes with a non-trivial source means it has a non-trivial normalization:
\begin{equation}
    \sqrt{Z_x(E)} = 1, \qquad \qquad
    \sqrt{Z_F(E)} = 1 + i \lambda E.
\end{equation}
This will play a role in the higher-point Green functions.
The 3-point ones are:
\begin{align}
    \langle \tilde{x}(E_1) \, \tilde{x}(E_2) \, \tilde{x}(E_3) \rangle_S
    &=
    \;
    \begin{tikzpicture}[baseline=-2pt]
        \draw (0, 0) -- (-0.5, 0);
        \draw (0, 0) -- (0.25, 0.44);
        \draw (0, 0) -- (0.25, -0.44);
        \draw[fill=black] (-0.5, 0) circle (2pt);
        \draw[fill=black] (0.25, 0.44) circle (2pt);
        \draw[fill=black] (0.25, -0.44) circle (2pt);
    \end{tikzpicture}
    \;
    =
    g \, \frac{1}{E^2 - \omega^2} 
    \frac{1}{E_1^2 - \omega^2} \frac{1}{E_2^2 - \omega^2}
    \delta(E_1 + E_2 + E_3),
    \\
    \langle \tilde{F}(E_1) \, \tilde{F}(E_2) \, \tilde{F}(E_3) \rangle_S
    &= 
    \;
    \begin{tikzpicture}[baseline=-2pt]
        \draw (0, 0) -- (-0.5, 0);
        \draw (0, 0) -- (0.25, 0.44);
        \draw (0, 0) -- (0.25, -0.44);
        \draw[fill=white] (-0.5, 0) circle (2pt);
        \draw[fill=white] (0.25, 0.44) circle (2pt);
        \draw[fill=white] (0.25, -0.44) circle (2pt);
    \end{tikzpicture}
    \; + \;
    \begin{tikzpicture}[baseline=-2pt]
        \draw (0.25, 0.44) -- (-0.5, 0);
        \draw (-0.5, 0) -- (0.25, -0.44);
        \draw[fill=white] (-0.5, 0) circle (2pt);
        \draw[fill=white] (0.25, 0.44) circle (2pt);
        \draw[fill=white] (0.25, -0.44) circle (2pt);
    \end{tikzpicture}
    \; + \;
    \begin{tikzpicture}[baseline=-2pt]
        \draw (-0.5, 0) -- (0.25, -0.44);
        \draw (0.25, 0.44) -- (0.25, -0.44);
        \draw[fill=white] (-0.5, 0) circle (2pt);
        \draw[fill=white] (0.25, 0.44) circle (2pt);
        \draw[fill=white] (0.25, -0.44) circle (2pt);
    \end{tikzpicture}
    \; + \;
    \begin{tikzpicture}[baseline=-2pt]
        \draw (0.25, 0.44) -- (0.25, -0.44);
        \draw (0.25, 0.44) -- (-0.5, 0);
        \draw[fill=white] (-0.5, 0) circle (2pt);
        \draw[fill=white] (0.25, 0.44) circle (2pt);
        \draw[fill=white] (0.25, -0.44) circle (2pt);
    \end{tikzpicture}
    \\
    =
    \delta(E_1 + E_2 + E_3) \, \Bigg[
    g
    &\frac{1 + i \lambda E_1}{E_1^2 - \omega^2}
    \frac{1 + i \lambda E_2}{E_2^2 - \omega^2}
    \frac{1 + i \lambda E_3}{E_3^2 - \omega^2}
    \nonumber \\
    +& \lambda \frac{1 + i\lambda E_1}{E_1^2 - \omega^2}
    \frac{1 + i\lambda E_2}{E_2^2 - \omega^2}
    + \lambda \frac{1 + i\lambda E_2}{E_2^2 - \omega^2}
    \frac{1 + i\lambda E_3}{E_3^2 - \omega^2}
    + \lambda \frac{1 + i\lambda E_1}{E_1^2 - \omega^2}
    \frac{1 + i\lambda E_3}{E_3^2 - \omega^2}
    \Bigg].
    \label{eq:example-3-point}
\end{align}
The two correlation functions are, in general, different. However,
as dictated by theorem~\ref{th:lsz-2}, they agree on-shell.
That is, in the limit $E_1, E_2 \to \omega$, we have the asymptotic relation
\begin{equation}
    \frac{\langle \tilde{x}(E_1) \, \tilde{x}(E_2) \, \tilde{x}(E_3) \rangle_S}{\sqrt{Z_x(E_1) Z_x(E_2) Z_x(E_3)}}
    \sim
    \frac{\delta(E_1 + E_2 + E_3) \mathcal{A}(E_1, E_2, E_3)}{(E_1^2 - \omega^2) (E_2^2 - \omega^2) (E_3^2 - \omega^2)}
    \sim
    \frac{\langle \tilde{F}(E_1) \, \tilde{F}(E_2) \, \tilde{F}(E_3) \rangle_S}{\sqrt{Z_F(E_1) Z_F(E_2) Z_F(E_3)}},
\end{equation}
where $\mathcal{A}(E_1, E_2, E_3) = g$.
It may be noted that the Dirac delta makes both functions vanish on-shell unless $\omega = 0$, making this relation trivial for $\omega > 0$.
However, after stripping the delta and normalizing with $Z_x$ and $Z_F$, the positions and residues at their on-shell poles still match for all $\omega$.

The diagrammatic approach gives an intuitive picture of how the classical LSZ theorem works.
Two theories that differ only on the source terms will have the same Feynman rules for internal lines and vertices.
The only difference between them comes from external vertices.
There can be two types: those with a single leg and those with multiple legs.
The latter is irrelevant in the on-shell limit because they will give contributions with fewer poles than necessary, as in the second line of Eq.~\eqref{eq:example-3-point}.
The contributions of the single-legged external vertices amount to a factor of $\sqrt{Z}$ per vertex.

\section{Issues with non-invertible field redefinitions}
\label{sec:non-invertible}

In Section~\ref{sec:interpolating}, we have assumed that the field redefinition functional $F[\varphi]$ is an interpolating field.
If this is not the case, $F$ will not be a one-to-one mapping of the field space itself, and physics will not be preserved by the redefinition.
This section studies how this happens and presents potential procedures to (partially) fix it.
We divide the discussion into two parts.
Firstly, we focus on field redefinitions that change the EOM, leading to new solutions, some of which can be eliminated through boundary conditions.
Secondly, we study the case in which a field redefinition generates a new EOM equivalent locally to the original one but changes the field space so that the solutions may differ globally.

\subsection{Non-perturbative redefinitions with derivatives}
\label{sec:with-derivatives}

Consider a local non-perturbative field redefinition, which may be written as
\begin{equation}
    \phi^x = F^x[\varphi]
    = f(\varphi(x),\partial\varphi(x), \cdots, \partial^N\varphi(x)),
\end{equation}
where $f$ is an ordinary function of $N+1$ variables.
$F$ will not be an interpolating field if a non-trivial field configuration $\check{\varphi}$ exists such that some non-trivial $\ psi$ solves the following differential equation:
\begin{equation}
    0 = \left.
        \frac{\delta F^x}{\delta \varphi^y}
    \right|_{\check{\varphi}} \psi^y
    \qquad \iff \qquad
    0 = \sum_{n = 0}^N
    \left.\frac{\partial f(x)}{\partial (\partial^n \varphi)}\right|_{\check{\varphi}}
    \partial^n \psi(x),
\end{equation}
where we have used the shorthand notation
\begin{equation}
    \partial^n \psi \equiv
    \partial_{\mu_1} \cdots \partial_{\mu_n} \psi,
    \qquad \qquad
    \left.
        \frac{\partial f(x)}{\partial(\partial^n \varphi)} 
    \right|_{\check{\varphi}} 
    \equiv 
    \frac{\partial f}{\partial(\partial^n\varphi)}
    (\check{\varphi}(x), \partial\check{\varphi}(x), \cdots, \partial^N \check{\varphi}(x)).
    \label{eq:F-singular}
\end{equation}
Additionally, contraction over the spacetime indices $\mu_j$ is understood.
Clearly, the functional $F$ is not invertible at $\check{\varphi}$ because the operator $\delta F^x/\delta \varphi^y$ is singular at that point in field space.
Although $F$ is local, it is not an interpolating field.

It is sufficient to consider the case in which $J = 0$ for the purpose of this section.
The EOM generated by the new action $\hat{S}[\varphi] = S[F[\varphi]]$ is
\begin{equation}
    0
    = \left.\frac{\delta \hat{S}}{\delta\varphi^x}\right|_{\varphi_0}
    = 
    \left.\frac{\delta S}{\delta \phi^y}\right|_{F[\varphi_0]}
    \left.\frac{\delta F^y}{\delta \varphi^x}\right|_{\varphi_0}.
    \label{eq:singular-eom-F}
\end{equation}
In terms of $f$, this equation is
\begin{equation}
    0 =
    \; \sum_{n = 0}^N
    \left.
        \frac{\partial f(x)}{\partial (\partial^n \varphi)}
    \right|_{\varphi_0}
    \partial^n
    \left.\frac{\delta S}{\delta\phi(x)}\right|_{F[\varphi_0]}.
    \label{eq:singular-eom-f}
\end{equation}
This is just Eq.~\eqref{eq:F-singular} with $\check{\varphi} = \varphi_0$ and $\psi = \delta S/\delta \phi |_{F[\varphi_0]}$.
This equation has typically more solutions than the original EOM $\delta S/\delta \phi = 0$.
We classify these solutions into two types:
\begin{enumerate}
    \item Field configurations $\varphi_0$ such that $\phi_0 = F[\varphi_0]$ is a solution to the original EOM.
    That is,
    \begin{equation}
        \left.\frac{\delta S}{\delta \phi(x)}\right|_{F[\varphi_0]}
        = 0.
        \label{eq:redefined-solution-eom}
    \end{equation}
    Essentially, $\varphi_0$ gives rise to the same physics as $\phi_0$ in the original theory.
    A difference with the original theory is that now there might be several different solutions $\varphi_0$, $\varphi'_0$ that correspond to the same original field $\phi_0$.
    This is because $F$ is not one-to-one, so it is possible that $F[\varphi_0] = F[\varphi'_0]$ when $\varphi_0 \neq \varphi'_0$.

    \item Field configurations $\varphi_0$ that do not solve the original EOM,
    \begin{equation}
        \left.\frac{\delta S}{\delta \phi(x)}\right|_{F[\varphi_0]}
        \neq 0,
    \end{equation}
    for some $x$, but solve the new EOM.
    That is, they satisfy Eq.~\eqref{eq:singular-eom-F} or, equivalently, Eq.~\eqref{eq:singular-eom-f}.
    This is possible only because $\delta F/\delta\varphi$ is singular: there are non-trivial solutions to Eq.~\eqref{eq:F-singular}.
    $\varphi_0$ is not related to any solution $\phi_0$ of the original theory through $F$.
    Thus, it may generate new physics.
\end{enumerate}
The existence of more solutions in the redefined theory than in the original one can be understood by noticing that the redefined EOM has a higher degree.
Let $d_i$ and $e_i$ be the number of derivatives and fields, respectively, in each term in the action $S = \sum_i S_i$; and let $M$ be the number of derivatives of the term with the most derivatives in $F$.
Then, the degree of the original EOM is
\begin{equation}
   \deg({\rm{EOM}})= \deg(S) = \max_i d_i \equiv D,
\end{equation}
while the degree of the redefined EOM is
\begin{equation}
    \deg({\rm E\widehat{O}M})=\deg(\hat{S}) = \max_i (d_i + e_i M) \leq D + E M, 
\end{equation}
where $E \equiv \max_i e_i$.
For $N > 0$, we always have $\deg(\hat{S}) > \deg(S)$.
If the terms in the action are at least quadratic in the fields ($e_i \geq 2$), there is a better lower bound: $\deg(\hat{S}) \geq D + 2 M$.

A specific solution of the original theory is usually determined by a set of $\deg(S)$ boundary conditions.
The natural procedure to do the same in the redefined theory is to perform the redefinition in the original boundary conditions to obtain boundary conditions for the redefined field $\varphi$.
The number of boundary conditions generated this way is $\deg(S) < \deg(\hat{S})$, which is typically not enough to select a single solution in the redefined EOM.

Nevertheless, one can eliminate the unphysical solutions (of the second type in the classification above) by imposing additional boundary conditions.\footnote{More precisely, in the following, we will have in mind initial conditions at a time $t$.}
To see how, note first that Eq.~\eqref{eq:F-singular} has degree $N$.
Thus, $N$ boundary conditions are generally required to determine a solution uniquely.
In particular, one can fix $\psi(x) \equiv 0$ by setting $\partial^n\psi(x) = 0$ (for all $n \leq N$) at the boundary.
As we know, the EOM~\eqref{eq:singular-eom-f} is just Eq.~\eqref{eq:F-singular} for $\psi = \delta S/\delta \phi$.
A set of conditions that eliminates the unphysical solutions is then
\begin{equation}
    \partial^{n}\frac{\delta S}{\delta\phi(x)}\bigg|_{F[\varphi_0]}
    = 0,
    \label{eq:boundary1}
\end{equation}
for all $n < N$ and $x$ in the boundary surface.
This is, however, additional information that needs to be provided together with the redefined action $\hat{S}$ and the redefined boundary conditions.

To provide a more concrete example of this discussion, we consider an action $S$ that is quadratic in the field and an $F$ that is linear.
Then, counting the number of solutions of each kind is particularly clear.
We have $M = N$, and
\begin{equation}
    \deg(\hat{S}) = \deg(S) + 2 N,
    \label{eq:quadratic-linear-degree}
\end{equation}
so the number of extra solutions is $2N$.
Half are of the first type, leaving the physics unchanged, and the other half are of the unphysical second type.
This can be seen by noticing that the degree of Eq.~\eqref{eq:redefined-solution-eom} is $\deg(S) + N$, while the degree of Eq.~\eqref{eq:F-singular} is $N$.

Let us now briefly return to the harmonic oscillator example from Section~\ref{sec:harmonic-oscillator}.
We consider the same field redefinition $x = F[y] = y + \lambda \dot{y}$, but we do not assume that $\lambda$ is a small parameter, so we allow solutions to depend on it non-perturbatively.
In this setup, the conditions for Eq.~\eqref{eq:quadratic-linear-degree} are satisfied, with $\deg(S) = 2$ and $N = 1$.
Thus, we know there should be 2 independent solutions for the original theory and 4 for the redefined one, with 3 being of the first type and 1 of the second type.
Indeed, the original EOM and its 2 solutions are:
\begin{equation}
    \ddot{x} + \omega^2 x = 0,
    \qquad \qquad
    x_\pm(t) = e^{\pm i \omega t}.
\end{equation}
The redefined EOM is Eq.~\eqref{eq:mechanical-redefined-eom}, which we reproduce here for completeness:
\begin{equation}
    (1 - \lambda^2 \partial_t^2) (\ddot{y} + \omega^2 y) = 0.
\end{equation}
Its 4 independent solutions are:
\begin{equation}
    y_\pm(t) = e^{\pm i \omega t},
    \qquad \qquad
    z_\pm(t) = e^{\pm t / \lambda}.
\end{equation}
The 3 solutions of the first type are $y_\pm$ and $y'_-$.
Indeed, it can be checked that $F[y_\pm] = (1 \pm i\lambda \omega) x_\pm$ and $F[y'_-] = 0$, all of which are solutions to the original EOM.
The only solution of the second type is $y'_+$.
It gives $F[y'_+] = 2 y'_+$, which is not a solution to the original EOM.
Finally, Eq.~\eqref{eq:boundary1} should eliminate $y'_+$ but not the others.
Specializing it for the present case, we get
\begin{equation}
    \left.\left\{
        \lambda \dddot{y}
        + \omega^2 (y + \lambda \dot{y})
        + \ddot{y}
    \right\}\right|_{t = 0} = 0.
\end{equation}
The solutions $y_\pm$ and $z_-$ satisfy this condition, but, as expected, $z_+$ does not.

\subsection{Redefinitions that do not preserve the field space}
\label{sec:solitons}

Let us switch our focus to redefinitions that are not one-to-one but do not change the EOM in the sense that the solutions to the redefined theory correspond locally (via $F$) to solutions of the original theory, except at isolated points.
We study this possibility through a concrete example:
a free real scalar field in $1 + 1$ dimensions with mass 1, whose action is
\begin{equation}
    S[\phi] = \frac{1}{2} \int d^2 x \, \left[
        (\partial_\mu \phi)^2 - \phi^2
    \right]
\end{equation}
We consider a field redefinition without derivatives:\footnote{We remark that the field variable is dimensionless in two dimensions. Therefore, the field redefinition does not necessarily contain a dimensionful redefinition parameter. That said, this is not crucial for the observations made in the following.}
\begin{equation}
    \phi = F[\varphi] = \varphi + \varphi^2.
\end{equation}
The functional $F$ is not one-to-one; thus, it cannot be an interpolating field.
This is easy to see since for $\phi > -1/4$, there are two values of $\varphi$ such that $F[\varphi] = \phi$.
They are given by
\begin{equation}
    \varphi = G_{\pm}[\phi] \equiv \frac{-1 \pm \sqrt{1 + 4 \phi}}{2}.
\end{equation}
For $\phi = -1/4$, the inverse is unique: $-1/4 = G_\pm[-1/2]$.
For $\phi < -1/4$ no real $\varphi$ such that $F[\varphi] = \phi$ exists.
Thus, all the solutions of the original theory in which $\phi(x) < -1/4$ at some point $x$ will be missing in the redefined theory.

The redefinition can nevertheless be made one-to-one in a restricted domain for fields in the ranges $\phi \geq -1/4$ and $\varphi \geq -1/2$, for example.
However, this requires providing additional information and the redefined action to recover the selected portion of the original theory.
We will concentrate here on an alternative possibility: given the redefined action only, does one only get solutions for it that map to solutions of the original theory, or do new solutions (and thus new physics) arise?

After the redefinition, the action is 
\begin{equation}
    \hat{S}[\varphi]
    = S[F[\varphi]]
    = \frac{1}{2} \int d^2x
    \left[
        (1 + 2 \varphi)^2 (\partial_\mu \varphi)^2
        - \varphi^2 (1 + \varphi)^2
    \right].
    \label{eq:solitons-redefined-action}
\end{equation}
It is immediately clear that the potential for the redefined theory has two degenerate minima at $\varphi = 0$ and $\varphi = -1$.
Typically, theories with degenerate minima (and a finite barrier separating them) contain solitons connecting them.
We will see that this is the case for the redefined theory.
The same is not true for the original theory, which has a single minimum at $\phi = 0$.

We will thus study the non-trivial finite-energy static solutions, called solitons, of the redefined theory and compare them with the ones for the original theory.
Let $x = (t, z)$.
Then, the energy functionals for static field configurations in both theories are
\begin{align}
    E[\phi] &= \frac{1}{2} \int dz \, 
    \left[(\partial_z \phi)^2 + \phi^2\right],
    \\
    \hat{E}[\varphi] &= \frac{1}{2} \int dz 
    \left[
        (1 + 2 \varphi)^2 (\partial_z \varphi)^2
        + \varphi^2 (1 + \varphi)^2
    \right].
\end{align}

For a static $\phi$ to have finite energy, it must be true that $\phi \to 0$ when $z \to \pm\infty$.
Similarly, a finite-energy $\varphi$ must be such that $\varphi \to -1/2 \pm 1/2$ as $z \to \pm\infty$.

The original theory only has two independent static solutions, both of which have \emph{infinite} energy:
\begin{equation}
    \phi_\pm(z) = -\frac{e^{\pm z}}{4}.
\end{equation}
Here, we have chosen a normalization that will be convenient for the following discussion.
Using $\phi_\pm$, we can construct a soliton in the redefined theory:
\begin{equation}
    \varphi_s(z) = \left\{\begin{array}{ll}
        G_-[\phi_-](z) & \text{if } z \leq 0, \\
        G_+[\phi_+](z) & \text{if } 0 < z.
    \end{array}\right.
\end{equation}
We display this solution in Fig.~\ref{fig:solitons}.
We show that this is indeed a static solution of the redefined theory in Appendix~\ref{sec:solitons-proof}.
It can also be directly checked that the energy of $\varphi_s$ is finite.

\begin{figure}
    \centering
    \includegraphics[width=0.9\linewidth]{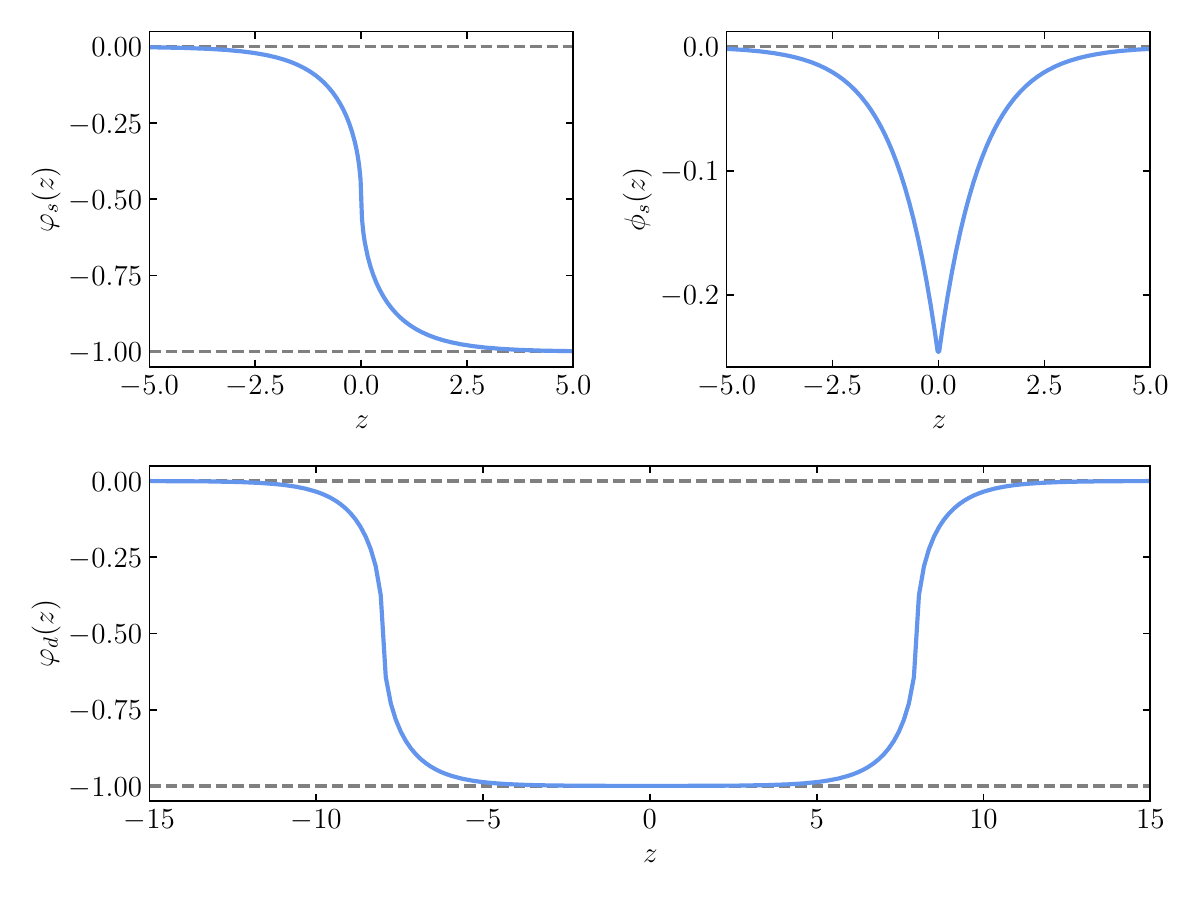}
    \caption{Top left: soliton solution of the redefined theory $\varphi_s$. Top right: field configuration $\phi_s = F[\varphi_s]$, which is not a solution of the original theory. Bottom: soliton-antisoliton solution $\varphi_d$ with the same boundary conditions as the trivial solution.}
    \label{fig:solitons}
\end{figure}

The physics of the redefined theory is then clearly different from the original one since it contains solitons, while the original one does not.
Remarkably, transforming $\varphi_s$ back into the original field a $\phi_s = F[\varphi_s]$ (shown in Fig.~\ref{fig:solitons}) does not generate a solution to the original one, as argued in Appendix~\ref{sec:solitons-proof}. Indeed, while $\phi_s$ is locally a solution everywhere in the domain $z \neq 0$, it does not solve the EOM at $z = 0$.

One might try to eliminate the extra solutions by means of boundary conditions, as in Section~\ref{sec:with-derivatives}.
For example, one could impose that $\varphi \to 0$ in both directions $x \to \pm \infty$, which restricts the solutions to those that look like the trivial one at infinity.
This, however, does not work because soliton-antisoliton solutions with this property can be constructed:
\begin{equation}
    \varphi_d(z) = \varphi_s\left(\frac{d}{2} - z\right)
    + \varphi_s\left(\frac{d}{2} + z\right)
    + 1.
\end{equation}
This is an approximate finite-energy static solution when the distance $d$ between the solitons is large.
We display it in Fig.~\ref{fig:solitons}.
Again, no analogue for this type of solution exists in the original theory.
This means the two theories cannot be rendered equivalent through boundary conditions.

\section{Conclusions and brief outlook to the quantum theory}
\label{sec:conclusions}

We have examined the properties and conditions for the applicability of field redefinitions in field theories, focusing on the classical case, which already contains several features such transformations have in quantum field theories.
One of the central results we have presented is a classical analogue for the LSZ theorem, using only purely classical concepts.
This result is connected to the semi-classical approximation of the quantum LSZ theorem.

Any field redefinition preserves the physics if it is a one-to-one mapping and depends locally on the field.
We have discussed two ways this statement can be made more concrete.
First, the physics of a given field theory should be independent of the parametrization used for the fields.
Indeed, as we have seen explicitly, the solutions to the original and the redefined EOM correspond one to one.
However, one should be careful when comparing observables before and after the redefinition.
For them to be preserved in general, they must all be redefined together with the action.
The second sense in which physics is invariant is a stronger statement: there is a set of observables (the positions and residues at the on-shell poles of Green functions) whose value remains unchanged under these transformations without any additional steps.
This is a consequence of the LSZ theorem, and focusing on this smaller set of essential observables that crucially contain cross-sections simplifies the study of field theories.
We have illustrated these results using two toy mechanical examples.

We have also investigated issues when the redefinition is not a one-to-one mapping.
Some solutions to the EOM might disappear in this case.
New solutions may also appear. These solutions are unphysical from the point of view of the original theory but can correspond to new phenomena contained in the new theory. 
Some can be eliminated by providing additional boundary conditions, which are additional information on top of the redefined version of the action and the original boundary conditions.
Other spurious solutions survive this type of procedure.
We have provided an example of this in which the new solutions are solitons.
When studying non-perturbative solutions, performing only valid redefinitions in the sense we have discussed is crucial.

The insights gained from this classical analysis provide a foundation for further exploration of quantum field theories.
The classical results we have presented apply to quantum systems at the tree level through the connection between the classical Green functions and the semi-classical approximation of vacuum expectation values of operator products discussed in Section~\ref{sec:quantum}.
They can also be extended to incorporate quantum effects, but additional issues arise.
This includes the impact of field redefinitions on the Jacobian factor in the path integral, their interaction with renormalisation and anomalies, and the effects of the field space domain in the path integration. We leave those interesting aspects for further work.

\section*{Acknowledgments}

JCC is supported by MICIU/AEI/10.13039/501100011033 and ERDF/EU (grants PID2022-139466NB-C22 and PID2021-128396NB-I00) and by the Ram\'on y Cajal program (grant RYC2021-030842-I). JJ gratefully acknowledges support by an IPPP DIVA fellowship, a Marsilius fellowship of Heidelberg University and the EU via ITN HIDDEN (No 860881).

\appendix

\section{Proofs of the classical LSZ theorems}
\label{sec:proofs}

\subsection{Theorem~\ref{th:lsz-1}}

First, let us study the properties of the solution $\phi_J$ to the EOM~\eqref{eq:J-eom} as a power series in $J$.
Expanding the left-hand side of this equation in powers of $J$ and requiring that each coefficient in the series vanishes, one gets an infinite set of equations of the form
\begin{align}
    0
    &=
    \left\{
    \frac{\delta^n}{\delta \tilde{J}_{p_1} \cdots \delta \tilde{J}_{p_n}}
    \left(\left.
    \frac{\delta S_J}{\delta \tilde{\phi}^p}
    \right|_{\phi_J}\right)
    \right\}_{J = 0}
    \\
    &=
    \left\{
    \frac{\delta^n}{\delta \tilde{J}_{p_1} \cdots \delta \tilde{J}_{p_n}}
    \left(\left.
    \frac{\delta S}{\delta \tilde{\phi}^p}
    \right|_{\phi_J}\right)
    +
    \sum_i
    \frac{\delta^{n-1}}{\delta \tilde{J}_{p_1} \cdots \tilde{J}_{p_{i-1}} \tilde{J}_{p_{i+1}} \delta \tilde{J}_{p_n}}
    \left(\left.
    \frac{\delta F^{p_i}}{\delta \tilde{\phi}^p}
    \right|_{\phi_J}\right)
    \right\}_{J = 0},
    \label{eq:proof-master-eq}
\end{align}
where we have used the DeWitt notation $\psi^p \equiv \psi_p \equiv \psi(p)$ for momenta in the same way we used it before for positions.
This set of equations can be solved iteratively.
We find it convenient to use the following simplified notation to do so.
Firstly, we drop the tilde $\tilde{\cdot}$ symbols and the $\cdot_S$ sub-indices for Green functions everywhere.
Secondly, we split
\begin{equation}
    S[\phi] = - \frac{1}{2} \phi^x \delta_{xy} (\square + m^2) \phi^y
    + S_{\rm int}[\phi].
\end{equation}
Finally, we define
\begin{equation}
    S_{p_1 \cdots p_n}
    \equiv
    \left.
        \frac{\delta^n S_{\rm int}}{\delta \phi^{p_1} \cdots \delta \phi^{p_n}}
    \right|_{\phi = 0},
    \qquad \qquad
    F^p_{p_1 \cdots p_n}
    \equiv
    \left.
        \frac{\delta^n F^p}{\delta \phi^{p_1} \cdots \delta \phi^{p_n}}
    \right|_{\phi = 0}.
\end{equation}
Importantly, note that the field derivatives are evaluated at $\phi=0$. This corresponds to evaluating in vacuum, as stated in the second point of Section~\ref{sec:lsz-results}.
Due to the locality of $S$ and $F$ these two objects are just finite-degree polynomials in the momenta.
Additionally, assumption~4 in Section~\ref{sec:lsz-results} implies that $S_{pq} = O(\lambda)$ if $F$ is $\lambda$-dependent (see also Eq 2.10), and $S_{pq} = 0$ otherwise.

\begin{figure}
    \centering    
    \includegraphics[width=0.95\textwidth]{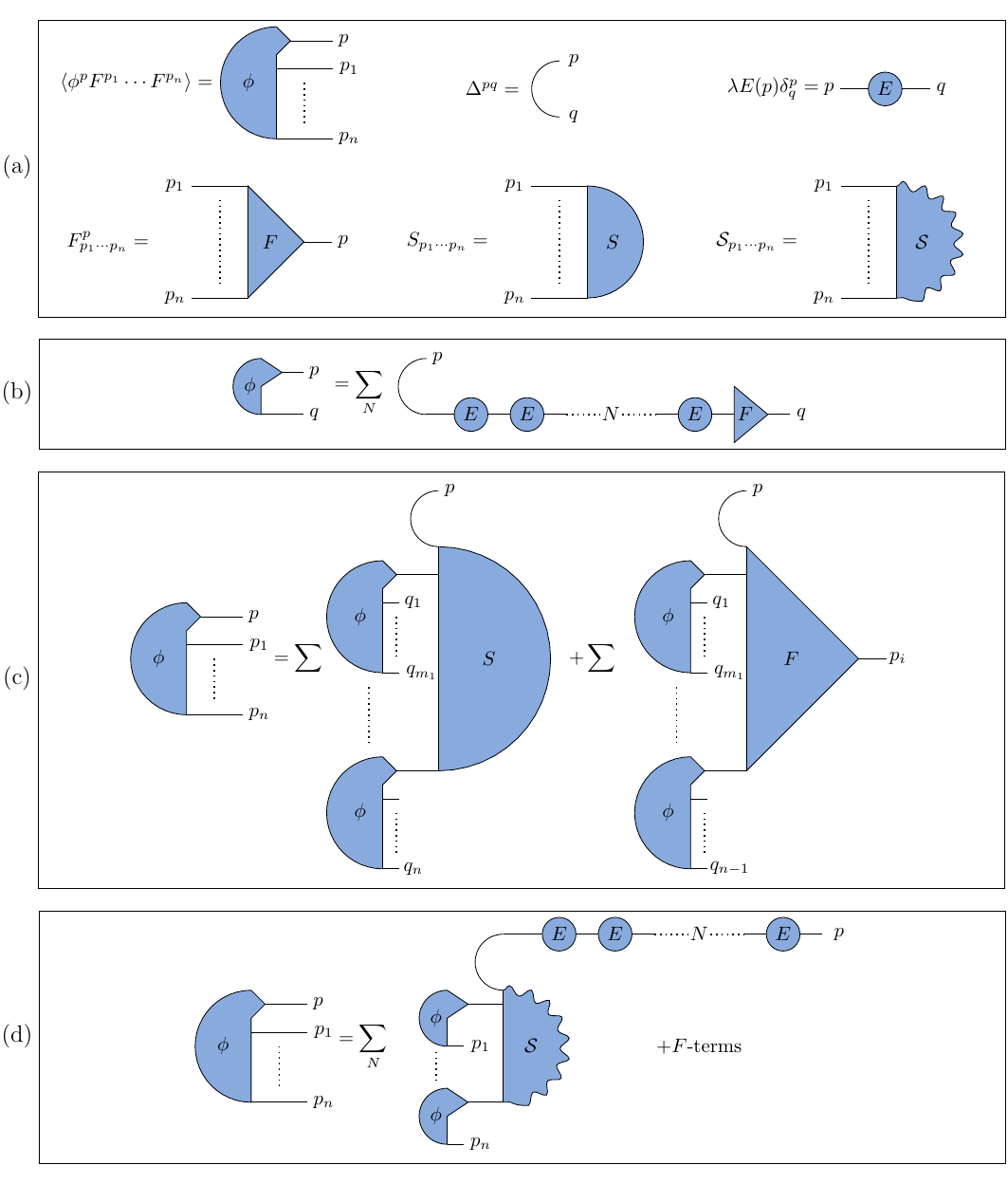}
    \caption{Diagrammatic representation of the equations in the proof of theorem~\ref{th:lsz-1}. Lines represent DeWitt momentum indices, with those pointing to the right (left) corresponding to upper (lower) ones. Panel~(a) provides the definitions of the basic building blocks. Panel~(b) gives the graphical representation of Eq.~\eqref{eq:proof-2-point-solution} for the 2-point functions. Panel~(c) represents Eq.~\eqref{eq:proof-master-schematic} relating the $n$-point function to $m$-point ones with $m < n$. Panel~(d) shows the diagrammatic version of Eq.~\eqref{eq:proof-field-solution} for the $n$-point function. A $\Delta$ line in all diagrams means a pole when the corresponding momentum goes on-shell. }
    \label{fig:proof-diagrammatic}
\end{figure}

We are now ready to focus on the iterative solution's starting point, the case $n = 1$.
Eq.~\eqref{eq:proof-master-eq} reduces to
\begin{equation}
    \langle\phi^p F^q\rangle
    =
    -\Delta^{p k} F^q_k 
    + \lambda E(p) \langle\phi^p F^q\rangle,
    \label{eq:proof-basic-case}
\end{equation}
where $\Delta^{pk} = \delta^{pk} / (p^2 - m^2)$ and $E(p)$ is defined in Eq.~\eqref{eq:classical-lsz-assumptions-free} and encapsulates the allowed perturbative momentum dependent corrections to the Klein Gordon equation a linear order in the fields.
If the field redefinition is non-perturbative, the second term on the right-hand side vanishes, and this equation provides an explicit formula for $\langle\phi^p F^q\rangle$.
Otherwise, the second term is $O(\lambda)$, and the equation can be used to obtain $\langle\phi^p F^q\rangle$ perturbatively in $\lambda$.
Explicitly, the solution is, in that case
\begin{equation}
    \langle\phi^p F^q\rangle
    =
    - \sum_{N = 0}^\infty \lambda^N E(p)^N \Delta^{p k} F^q_k.
    \label{eq:proof-2-point-solution}
\end{equation}
A diagrammatic representation of this equation is shown in Fig.~\ref{fig:proof-diagrammatic}.
At any rate, the solution has the desired structure for theorem~\ref{th:lsz-1}, with a delta function for $p$ and $q$ and a pole when they go on-shell.
The residue is guaranteed to be non-vanishing because $F$ is an interpolating field, which implies that $F^q_k \neq 0$.

The next step is to proceed by induction: we assume that theorem~\ref{th:lsz-1} is true for all the $m$-point functions $\langle\phi^p F^{p_1} \cdots F^{p_m}\rangle$ with $m < n$ and prove it for $n$-point function $\langle\phi^p F^{p_1} \cdots F^{p_n}\rangle$.
For $n>1$, Eq.~\eqref{eq:proof-master-eq} provides the kind of relation we need.
Schematically:
\begin{align}
    \langle \phi^p F^{p_1} \cdots F^{p_n}\rangle
    &=
    - \Delta^{pk}
    \Big[
    \sum S_{k k_1 k_2 \cdots } \langle\phi^{k_1} F^{q_1} \cdots F^{q_{m_1}}\rangle
    \langle\phi^{k_2} F^{q_{m_1 + 1}} \cdots F^{q_{m_1 + m_2}}\rangle \cdots
    \nonumber
    \\
    &\phantom{=-\Delta^{pk}\Big[}
    + \sum F^p_{k k_1 k_2\cdots} \langle\phi^{k_1} F^{q_1} \cdots F^{q_{m_1}}\rangle
    \langle\phi^{k_2} F^{q_{m_1 + 1}} \cdots F^{q_{m_1 + m_2}} \rangle \cdots 
    \Big],
    \label{eq:proof-master-schematic}
\end{align}
where the first sum is over all partitions $m_1 + m_2 + \cdots = n$ with $m_j > 0$, and all permutations of the momenta $(p_1, \cdots, p_n) \to (q_1, \cdots, q_n)$; while the second sum is over all $0 < i \leq n$, all partitions $m_1 + m_2 + \cdots = n - 1$, and all permutations $(p_1, \cdots, p_{i - 1}, p_{i + 1}, \cdots, p_n) \to (q_1, \cdots, q_{n-1})$.
The terms in the first sum are products of $m$-point functions with $m < n$.
By the induction assumption, they will have a pole in all the external momenta $p_j$ when they go on-shell.
The terms in the second sum are also products of $m$-point functions, but they will be missing the pole when the momentum $p_i$ goes on-shell, so they are irrelevant for the asymptotic behaviour in the on-shell limit.
This proves that the $n$-point function has the desired pole structure.
We present a diagrammatic version of this formula in Fig.~\ref{fig:proof-diagrammatic}.

A small caveat arises where $F$ is only perturbatively an interpolating field. The first sum has terms with $m = n$ in this case.
However, such terms will be $O(\lambda)$ instead of the rest, $O(\lambda^0)$.
One then needs to proceed as for Eq.~\eqref{eq:proof-basic-case} and solve the equation perturbatively in $\lambda$.
Then, the leading order has the correct pole structure.
The higher orders inherit this structure from the leading one.

To complete the proof for theorem~\ref{th:lsz-1}, it remains to show that the residue at the on-shell pole splits as displayed in Eq.~\eqref{eq:th-lsz-1}.
To see this, consider replacing the $m$-point functions in Eq.~\eqref{eq:proof-master-schematic} by the right-hand side of their own version of Eq.~\eqref{eq:proof-basic-case} repeatedly.
At the end of this process, the right-hand side of the equation will only contain 2-point functions $\langle\phi^p F^q\rangle$.
That is,
\begin{equation}
    \langle \phi^p F^{p_1} \cdots F^{p_n}\rangle
    =
    \Delta^{pq} \mathcal{S}_{q k_1 \cdots k_n}
    \langle\phi^{k_1} F^{p_1}\rangle
    \cdots
    \langle\phi^{k_n} F^{p_n}\rangle
    + \lambda E(p) \langle\phi^p F^{p_1} \cdots F^{p_n}\rangle
    + F\text{-terms}.
    \label{eq:proof-master-simplified}
\end{equation}
where $F$-terms denotes terms in which $F^p_{p_1\cdots}$ appears explicitly, and thus have less poles than the first term, and $\mathcal{S}_{q k_1 \cdots k_n}$ is a function of the momenta, constructed as a product of the functions $\Delta^{q_1 q_2}$ and $S_{q_1 q_2 \cdots}$, where the $q_j$ are sums of subsets of the $k_j$.
The perturbative (in $\lambda$) solution to this equation is
\begin{equation}
    \langle \phi^p F^{p_1} \cdots F^{p_n}\rangle
    =
    \sum_{N = 0}^\infty 
    \lambda^N E(p)^N \Delta^{pq}
    \mathcal{S}_{q k_1 \cdots k_n}
    \langle\phi^{k_1} F^{p_1}\rangle
    \cdots
    \langle\phi^{k_n} F^{p_n}\rangle
    + F\text{-terms}.
    \label{eq:proof-field-solution}
\end{equation}
The diagram for this equation is displayed in Fig.~\ref{fig:proof-diagrammatic}.
If $F$ is $\lambda$-independent, only the $N = 0$ term appears.
Conservation of momentum dictates that $\mathcal{S}_{q k_1 \cdots k_n}$ has a global $\delta^d(p + k_1 + \cdots + k_n)$ factor, so we can write
\begin{equation}
    \mathcal{S}_{k_1 \cdots k_n} = 
    \delta^d(k_1 + \cdots + k_n)
    \mathcal{A}_{k_1 \cdots k_n}.
    \label{eq:proof-A-definition}
\end{equation}
$\mathcal{A}_{k_1 \cdots k_n}$ is a function of momenta that depends only on the action $S$ and is analytic everywhere except when sums of momenta go on-shell.

Beyond that we can see from Eq.~\eqref{eq:proof-field-solution} that the new pole in $p$ always comes together with the same pre-factor that can be associated with $Z$ and reflects the only $\lambda$-dependence of the pole.
Using theorem~\ref{th:lsz-1} to define the $Z$ factors for the 2-point functions, our solution for the $n$-point one satisfies Eq.~\eqref{eq:th-lsz-1}, finishing the proof.

\subsection{Theorem~\ref{th:lsz-2}}

We first notice the following relation:
\begin{equation}
    \frac{\delta W}{\delta J_x}
    = 
    \left.\frac{\delta S_J}{\delta \phi^y}\right|_{\phi_J}
    \frac{\delta \phi^y_J}{\delta J_x}
    + \left.\frac{\delta S_J}{\delta J_x}\right|_{\phi_J}
    = F^x[\phi_J],
\end{equation}
where the first equality comes from the definition of $W[J]$, and the second one from the fact that $\phi_J$ is a solution to the EOM and $\delta S_J/\delta J = F$.
The $n$-point function can then be computed as
\begin{equation}
    \langle F^{p_1} \cdots F^{p_n}\rangle
    =
    \left.
        \frac{\delta^{n-1}}{J_{p_2} \cdots J_{p_n}} F^{p_1}[\phi_J]
    \right|_{J = 0}
    =
    F^{p_1}_q \langle\phi^{q} F^{p_2} \cdots F^{p_n}\rangle
    + F^{(2)}\text{-terms},
    \label{eq:proof-Green-field-relation}
\end{equation}
where $F^{(2)}$-terms denotes terms with a factor of $F^{p_1}_{q_1\cdots q_n}$ with $n\geq 2$, and thus several factors of $\langle \phi^{q_1} F^{p_j} \cdots\rangle$.
The first term on the right-hand side has a pole when $p_1^2 \to m^2$, but none of the $F^{(2)}$-terms does, so they are irrelevant for the asymptotic behaviour of the $n$-point function in the on-shell limit.
Substituting the solution Eq.~\eqref{eq:proof-field-solution} into Eq.~\eqref{eq:proof-Green-field-relation} gives
\begin{equation}
    \langle F^{p_1} \cdots F^{p_n}\rangle
    =
    \mathcal{S}_{k_1 \cdots k_n}
    \langle\phi^{k_1} F^{p_1}\rangle
    \cdots
    \langle\phi^{k_n} F^{p_n}\rangle
    + F\text{-terms}.
    \label{eq:proof-field-solution2}
\end{equation}
Applying theorem~\ref{th:lsz-1} to the 2-point functions that appear here, together with Eq.~\eqref{eq:proof-A-definition}, proves theorem~\ref{th:lsz-2}.

\section{Spurious soliton solutions}
\label{sec:solitons-proof}

We will show here that the field configuration $\varphi_s$ defined in Section~\ref{sec:solitons} is a solution to the redefined theory $\hat{S}$, while its cousin $\phi_s = F[\varphi_s]$ is not a solution of the original theory $S$.

Let us first give a precise definition of what counts as a solution.
Throughout the paper, we have interpreted the EOM $\delta S/\delta \phi$ as a differential equation whose solutions must be sufficiently smooth for the derivatives appearing in $\delta S/\delta \phi$ to be well-defined.
We provide here a slightly more general definition.
A solution to the theory defined by $S$ is function $\phi$ that extremizes $S$. That is,
\begin{equation}
    \frac{\delta S}{\delta \phi^x} \psi^x = 0,
    \label{eq:eom-distributions}
\end{equation}
for all test functions $\psi$.
If $\phi$ is assumed to be smooth, this is equivalent to interpreting $\delta S/\delta \phi$ as a differential equation over $\phi$.
However, one may allow $\phi$ to be singular at certain points and interpret $\delta S/\delta \phi$ as a distribution acting on $\psi$.
This useful for dealing with $\phi_s$ and $\varphi_s$, since both of them are singular at $z = 0$: the spatial derivative $\partial_z \phi_s$ is discontinuous there, while $\partial_z \varphi_s$ diverges.

This definition is mathematically valid, but it may raise questions regarding its physical interpretation.
To address this, one can always regularize $\phi_s$ and $\varphi_s$ by deforming them in a small region near $z = 0$ to make them smooth.
$\phi_s$ and $\varphi_s$ can then be understood as a coarse-grained approximation to the smooth solutions.
For simplicity, we will not discuss this regularization in detail here and instead focus on the singular functions directly.

Substituting the actions $S$ and $\hat{S}$ of Section~\ref{sec:solitons} into Eq.~\eqref{eq:eom-distributions}, and focussing on the case of static solutions, the corresponding EOMs read
\begin{gather}
    \int dz  \left\{
        \partial_z^2 \phi - \phi
    \right\} \psi = 0,
    \label{eq:eom-distribution-phi}
    \\
    \int dz \left\{
        (1 + 2 \varphi)^2 \, \partial_z^2 \varphi
        + 2 (1 + 2 \varphi) (\partial_z \varphi)^2
        - \varphi (1 + \varphi) (1 + 2 \varphi)
    \right\} \psi = 0.
    \label{eq:eom-distribution-varphi}
\end{gather}
For any $\psi$ whose support does not contain $z = 0$, only the smooth region of $\phi$ and $\varphi$ is relevant. Then, both equations are equivalent to the vanishing of the corresponding differential equations in brackets.
The equation for $\phi$ clearly satisfied by $\phi_s$, because $\phi_\pm$ are both smooth solutions to their EOM.
Additionally, in this region, the EOM for $\varphi$ is equivalent to the EOM for $F[\varphi]$ because in it $\varphi \neq -1/2$, making $F$ locally one-to-one.
Thus $\varphi_s$ is also a solution there.

It only remains to see what happens when the support of $\psi$ contains $z = 0$.
It is sufficient to consider the limit when $\epsilon \to 0$ of a family of test functions $\psi_\epsilon$ defining a small window around the origin, as
\begin{equation}
    \psi_\epsilon(x) = \left\{\begin{array}{ll}
        1 & \text{if } -\epsilon \leq x \leq \epsilon \\
        0 & \text{otherwise}
    \end{array}\right..
\end{equation}
Eqs.~\eqref{eq:eom-distribution-phi} and~\eqref{eq:eom-distribution-varphi} are then equivalent to the vanishing of the following integrals
\begin{align}
    I_\epsilon[\phi] &=
    \frac{\delta S}{\delta \phi^x} \psi_\epsilon^x
    =
    \int_{-\epsilon}^\epsilon dz \left\{
        \partial_z^2 \phi - \phi
    \right\},
    \\
    \hat{I}_\epsilon[\varphi] &= 
    \frac{\delta \hat{S}}{\delta \varphi^x} \psi_\epsilon^x
    =
    \int_{-\epsilon}^\epsilon dz \left\{
        (1 + 2 \varphi)^2 \, \partial_z^2 \varphi
        + 2 (1 + 2 \varphi) (\partial_z \varphi)^2
        - \varphi (1 + \varphi) (1 + 2 \varphi)
    \right\},
\end{align}
when $\epsilon \to 0$.
We remark that, because of the singularities at $z = 0$, one cannot evaluate $I_\epsilon[\phi_s]$ and $\hat{I}_\epsilon[\varphi_s]$ by taking the expression for the integrand at $z \neq 0$ and performing the integral over it naively.
However, this can be done for the terms without derivatives in both equations since they are non-singular.

For the $\partial_z^2 \phi$ term, a valid manipulation, in the sense of distributions, is to apply the fundamental theorem of calculus, which gives
\begin{equation}
    \lim_{\epsilon \to 0} I_\epsilon[\phi_s] 
    =
    \lim_{\epsilon \to 0} [\partial_z \phi_s(\epsilon) - \partial_z \phi_s(-\epsilon)] = -1/2.
    \label{eq:phis-not-sol}
\end{equation}
This result can be obtained by noticing that $\partial_z^2 \phi$ contains a Dirac delta at $z = 0$, since the first derivative $\partial_z \phi$ has a finite jump.
Eq.~\eqref{eq:phis-not-sol} implies that Eq.~\eqref{eq:eom-distribution-phi} is not satisfied for $\phi = \phi_s$ and $\psi = \psi_\epsilon$, so $\phi_s$ is not a solution of the original theory.

Regarding the $(1 + 2\varphi)^2 \partial_z^2 \varphi$ term, another valid manipulation is to perform integration by parts.
This gives
\begin{equation}
    \lim_{\epsilon \to 0}\hat{I}_\epsilon[\varphi_s] = 
    \lim_{\epsilon \to 0}
    \left\{
    \left[
        (1 + 2\varphi_s)^2 \partial_z \varphi_s
    \right]_{-\epsilon}^{+\epsilon}
    - 2 \int_{-\epsilon}^\epsilon dz
    (1 + 2 \varphi_s) (\partial_z \varphi_s)^2
    \right\}.
\end{equation}
Now, because only small values of $z$ are involved, we may substitute the following small-$z$ approximation of $\varphi_s$:
\begin{equation}
    \varphi_s(z) = \frac{\operatorname{sign}(z) \sqrt{|z|} - 1}{2}
    + O(z^{3/2}).
\end{equation}
The first term then clearly vanishes, and we are left with
\begin{equation}
    \lim_{\epsilon \to 0}\hat{I}_\epsilon[\varphi_s] = 
    -\frac{1}{8} \lim_{\epsilon \to 0}
    \int_{-\epsilon}^\epsilon dz
    \frac{\operatorname{sign}(z)}{\sqrt{|z|}}
    = -\frac{1}{4} \lim_{\epsilon \to 0}
    \left[\sqrt{|z|}\right]_{-\epsilon}^{+\epsilon}
    = 0.
\end{equation}
Therefore, $\varphi_s$ is a solution of the redefined theory.

\bibliographystyle{utphys}
\bibliography{references.bib}
\end{document}